\documentclass[runningheads,a4paper]{llncs}

\usepackage[utf8x]{inputenc}
\usepackage{textgreek}
\usepackage{amssymb}
\usepackage{graphicx}
\usepackage{url}
\usepackage{alltt}
\usepackage{amsmath}
\usepackage{proof}
\usepackage{todonotes}

\newcommand{\abs}[4]{{#1}\, #2\! : \! #3.\, #4}
\newcommand{\lam}[2]{\lambda\, #1.\, #2}

\DeclareUnicodeCharacter{9666}{$\blacktriangleleft$}
\DeclareUnicodeCharacter{10132}{$\rightarrow$}
\DeclareUnicodeCharacter{9733}{$\star$}

\DeclareUnicodeCharacter{953}{$\iota$}       
\DeclareUnicodeCharacter{955}{\textlambda}
\DeclareUnicodeCharacter{923}{\textLambda}
\DeclareUnicodeCharacter{928}{\textPi}
\DeclareUnicodeCharacter{946}{\textbeta}
\DeclareUnicodeCharacter{183}{$\!\!\!$}
\DeclareUnicodeCharacter{10174}{$\rightarrow$}
\DeclareUnicodeCharacter{2081}{$_1$}
\DeclareUnicodeCharacter{2082}{$_2$}
\DeclareUnicodeCharacter{7522}{$_i$}

\urldef{\mailsa}\path|{name-lastname}@uiowa.edu|    
\newcommand{\keywords}[1]{\par\addvspace\baselineskip
\noindent\keywordname\enspace\ignorespaces#1}

\begin{document}

\mainmatter  

\title{Efficient Mendler-Style Lambda-Encodings in Cedille}

\authorrunning{Denis Firsov, Richard Blair, Aaron Stump}

%
%
\author{Denis Firsov, Richard Blair, and Aaron Stump}

\institute{
Department of Computer Science \\
The University of Iowa \\
Iowa City, IA, USA \\
\mailsa}

\maketitle

\begin{abstract}
It is common to model inductive datatypes as least fixed points of
functors. We show that within the Cedille type theory we can relax
functoriality constraints and generically derive an induction
principle for Mendler-style lambda-encoded inductive datatypes, which
arise as least fixed points of covariant schemes where the morphism
lifting is defined only on identities.  Additionally, we implement a
destructor for these lambda-encodings that runs in constant-time. As a
result, we can define lambda-encoded natural numbers with an induction
principle and a constant-time predecessor function so that the normal
form of a numeral requires only linear space. The paper also includes
several more advanced examples.


\end{abstract}

\section{Introduction}
It is widely known that inductive datatypes may be defined in pure
impredicative type theory. For example, Church encodings identify
each natural number \verb;n; with its iterator \verb;λ s. λ z. sⁿ z;.
The Church natural numbers can be typed in System~F by means of
impredicative polymorphism:
\begin{alltt}
cNat ◂ ★ = ∀ X : ★. (X ➔ X) ➔ X ➔ X.
\end{alltt}
The first objection to lambda-encodings is that it is provably
impossible to derive an induction principle  in second-order dependent
type theory~\cite{geuvers01}.  As a consequence, most languages come
with a built-in infrastructure for defining inductive datatypes and
their induction principles. Here are the
definitions of natural numbers in Agda and Coq:
\begin{alltt} 
data Nat : Set                   Inductive nat : Type :=
  zero : Nat                       | 0 : nat
  suc  : Nat ➔ Nat                | S : nat → nat.
\end{alltt}
Coq will automatically generate the induction principle for \verb;nat;,
and in Agda it can be derived by pattern matching and explicit
structural recursion.

Therefore, we can ask if it is possible to extend the Calculus of
Constructions with {\bf{typing constructs}} that make induction
derivable for lambda-encoded datatypes. Stump gave a positive answer
to this question by introducing the Calculus of Dependent Lambda
Eliminations (CDLE)~\cite{stump17a}. CDLE is a Curry-style Calculus of
Constructions extended with implicit products, intersection types, and
primitive heterogeneous equality. Stump proved that natural number
induction is derivable in this system for lambda-encoded natural
numbers. Later, we generalized this work by deriving induction for
lambda-encodings of inductive datatypes which arise as least fixed
points of functors~\cite{firsov18}. Moreover, we observed that the
proof of induction for Mendler-style lambda-encoding relied only on
the identity law of functors. In this paper, we exploit this
observation to define a new class of covariant schemes, which includes
functors, and induces a larger class of inductive datatypes supporting
derivable induction.

Another objection to lambda-encodings is their computational
inefficiency. For example, computing the predecessor of a Church
encoded Peano natural provably requires linear
time~\cite{parigot89}. The situation was improved by Parigot who
proposed a new lambda-encoding of numerals with a constant-time
predecessor, but the size of the number $n$ is exponential
$O(2^n)$~\cite{parigot88}.  Later, the situation was improved further
by the Stump-Fu encoding, which supports a constant-time predecessor
and reduces the size of the natural number $n$ to
$O(n^2)$~\cite{stump16}. In this paper, we show how to develop a
constant-time predecessor within CDLE for a Mendler-style
lambda-encoded naturals that are linear in space.

This paper makes the following technical contributions:
\begin{enumerate}
\item We introduce a new kind of parameterized scheme using
  {\bf{identity mappings}} (function lifting defined only
  on identities). We show that every functor has an associated
  identity mapping, but not vice versa.

\item We use a Mendler-style lambda-encoding to prove that every
  scheme with an identity mapping induces an inductive datatype. Additionally,
  we generically derive an induction principle for these datatypes.

\item We implement a generic constant-time destructor of Mendler-style
  lambda-encoded inductive datatypes.  To the best of our knowledge,
  we offer a first example of typed lambda-encoding of inductive
  datatypes with derivable induction and a constant-time destructor
  where normal forms of data require linear space.

\item We give several examples of concrete datatypes defined using our
  development. We start by giving a detailed description of lambda-encoded
  naturals with an induction principle and a constant-time predecessor
  function that only requires linear space to encode a numeral. We
  also give examples of infinitary datatypes.  Finally, we present an
  inductive datatype that arises as a least fixed point of a scheme
  that is not a functor, but has an identity mapping.

\end{enumerate}

\section{Background}
In this section, we briefly summarize the main features of Cedille's
type theory.  For full details on CDLE, including semantics and
soundness results, please see the previous
papers~\cite{stump17a,stump18}.  The main metatheoretic property
proved in the previous work is logical consistency: there are types
which are not inhabited.  CDLE is an extrinsic (i.e., Curry-style)
type theory, whose terms are exactly those of the pure untyped lambda
calculus (with no additional constants or constructs).  The
type-assignment system for CDLE is not subject-directed, and thus
cannot be used directly as a typing algorithm.  Indeed, since CDLE
includes Curry-style System F as a subsystem, type assignment is
undecidable~\cite{Wells99}.  To obtain a usable type theory, Cedille
thus has a system of annotations for terms, where the annotations
contain sufficient information to type terms algorithmically.  But
true to the extrinsic nature of the theory, these annotations play no
computational role.  Indeed, they are erased both during compilation
and before formal reasoning about terms within the type theory, in
particular by definitional equality (see Figure~1).

\begin{figure}
\centering
\minipage[b]{0.8\textwidth}
  \[
  \begin{array}{cc}
    \infer{\Gamma\vdash \abs{\Lambda}{x}{T'}{t} : \abs{\forall}{x}{T'}{T}}{\Gamma,x:T'\vdash t : T & \!\!x\not\in\textit{FV}(|t|)} & 
    \infer{\Gamma\vdash t\ -t' : [t'/x]T}{\Gamma\vdash t : \abs{\forall}{x}{T'}{T} & \Gamma\vdash t':T'} \\ \\

    \infer{\Gamma\vdash \beta : t \simeq t}{\Gamma\vdash t : T} &
    \infer{\Gamma\vdash \rho\ t'\ -\ t : [t_2/x]T}{\Gamma\vdash t' : t_1 \simeq t_2 & \Gamma \vdash t : [t_1/x]T} \\ \\

    \multicolumn{2}{l}{\infer{\Gamma\vdash [t_1,t_2\{p\}] : \abs{\iota}{x}{T}{T'}}{\Gamma\vdash t_1 : T & \Gamma\vdash t_2 : [t_1/x]T' & \Gamma\vdash p : |t_1| \simeq |t_2|}} \\ \\

    \infer{\Gamma\vdash t.1 : T}{\Gamma\vdash t : \abs{\iota}{x}{T}{T'}} &
    \infer{\Gamma\vdash t.2 : [t.1/x]T'}{\Gamma\vdash t : \abs{\iota}{x}{T}{T'}} \\ \\

  \end{array}
  \]
\endminipage\hfill 
\minipage[b]{0.20\textwidth}
\[
  \begin{array}{lll}
    |\abs{\Lambda}{x}{T}{t}| & = & |t| \\
    |t\ -t'| & = & |t| \\
    |\beta| & = & \lam{x}{x} \\
    |\rho\ t\ - \ t'| & = & |t'| \\
    |[t_1,t_2\{p\}]| & = & |t_1| \\
    |t.1| & = & |t| \\
    |t.2| & = & |t| 
  \end{array}
  \]
\endminipage
\caption{Introduction, elimination, and erasure rules for additional type constructs} 
\end{figure}
CDLE extends the (Curry-style) Calculus of Constructions (CC) with
primitive heterogeneous equality, intersection types, and implicit
products:
\begin{itemize}
\item \verb;t₁ ≃ t₂;, a heterogeneous equality type.  The terms
  \verb;t₁; and \verb;t₂; are required to be typed, but need not have
  the same type.  We introduce this with a constant \verb;β; which
  erases to \verb;λ x. x; (so our type-assignment system has no
  additional constants, as promised); \verb;β; proves \verb;t ≃ t; for
  any typeable term \verb;t;.  Combined with definitional equality,
  \verb;β; proves \verb;t₁ ≃ t₂; for any $\beta$-equal \verb;t₁; and
  \verb;t₂; whose free variables are all declared in the typing
  context.  We eliminate the equality type by rewriting, with a
  construct \verb;ρ t' - t;.  Suppose \verb;t'; proves \verb;t₁ ≃ t₂;
  and we synthesize a type \verb;T; for \verb;t;, where \verb;T; has
  several occurrences of terms definitionally equal to \verb;t₁;.  Then
  the type synthesized for \verb;ρ t' - t; is \verb;T; except with
  those occurrences replaced by \verb;t₂;.  Note that the types of the
  terms are not part of the equality type itself, nor does the
  elimination rule require that the types of the left-hand and right-hand
  sides are the same to do an elimination.

\item \verb;ι x : T. T';, the dependent intersection type of
  Kopylov~\cite{kopylov03}.  This is the type for terms \verb;t; which
  can be assigned both the type \verb;T; and the type \verb;[t/x]T';,
  the substitution instance of \verb;T'; by \verb;t;.  In the
  annotated language, we introduce a value of \verb;ι x : T. T'; by
  construct \verb;[ t, t' {p} ];, where \verb;t; has type \verb;T;
  (algorithmically), \verb;t'; has type \verb;[t/x]T';, and \verb;p;
  proves \verb;t ≃ t';.  There are also annotated constructs
  \verb;t.1; and \verb;t.2; to select either the \verb;T; or
  \verb;[t.1/x]T'; view of a term \verb;t; of type \verb;ι x : T. T';.

\item \verb;∀ x : T. T';, the implicit product type of
  Miquel~\cite{miquel01}.  This can be thought of as the type for
  functions which accept an erased input of type \verb;x : T;, and
  produce a result of type \verb;T';. There are term constructs
  \verb;Λ x. t; for introducing an implicit input \verb;x;, and
  \verb;t -t'; for instantiating such an input with \verb;t';. The
  implicit arguments exist just for purposes of typing so that they
  play no computational role and equational reasoning happens on terms
  from which the implicit arguments have been erased.
\end{itemize}
It is important to understand that the described constructs are erased
before the formal reasoning, according to the erasure rules in
Figure~1.

We have implemented CDLE in a tool called Cedille, which we have used
to typecheck the developments of this paper. The pre-release version
is here:
\vspace{.1cm}

\url{http://cs.uiowa.edu/~astump/cedille-prerelease.zip}

\vspace{.1cm}

\noindent The Cedille code accompanying this paper is here:

\vspace{.1cm}

\url{http://firsov.ee/efficient-lambda/itp2018-code.zip}
\section{Preliminaries} 
We skip the details of the lambda-encoded implementation of basic
datatypes like \verb;Unit;, \verb;Empty;, sums (\verb;X + Y;), and
dependent products (\verb;Σ x : X. Y x;), for which the usual
introduction and elimination rules are derivable in Cedille.

In this paper, we use syntactical simplifications to improve
readability. In particular, we hide the type arguments in the cases
when they are unambiguous. For example, if \verb;x : X; and
\verb;y : Y; then we wrtie \verb;pair x y; instead of fully
type-annotated \verb;pair X Y x y;. The current version of Cedille
requires fully annotated terms.

\subsection{Multiple Types of Terms}
CDLE's dependent intersection types allow judgementally
equal values to be intersected. Given \verb;x : X;, \verb;y : Y x;,
and a proof \verb;p; of \verb;x ≃ y;, we can introduce an intersection value
\verb;v := [ x, y {p} ]; of type \verb;ι x : X. Y x;. Every
intersection has two ``views'': the first view \verb;v.1; has type
\verb;X; and the second view \verb;v.2; has type \verb;Y x;.  The term
\verb;[ x, y {p} ]; erases to \verb;x; according to the erasure
rules in Figure~1.  This allows us to see \verb;x; as having two
distinct types, namely \verb;X; and \verb;Y x;:
\begin{alltt}
subst ◂ ∀ X: ★. ∀ Y: X ➔ ★. Π x: X. ∀ y: Y x. ∀ p: x ≃ y. Y x 
 = Λ X. Λ Y. λ x. Λ y. Λ p. [ x, y \{p\} ].2.
\end{alltt}
(\verb;Π x : X. T; is usual ``explicit'' dependent function space;)
Indeed, the definition of \verb;subst; erases to term \verb;λ x. x;.
Hence, \verb;subst x -y -p; beta-reduces to \verb;x; and has type
\verb;Y x; (dash denotes the application of implicitly quantified arguments).

\subsection{Identity Functions}
In our setting, it is possible to implement a function of type
\verb;X ➔ Y; so that it erases to term \verb;λ x. x; where \verb;X; is
different from \verb;Y;. The simplest example is the first (or second)
``view'' from an intersection value:
\begin{alltt}
view1 ◂ ∀ X : ★. ∀ Y : X ➔ ★. (ι x : X. Y x) ➔ X 
 = Λ X. Λ Y. λ x. x.1.
\end{alltt}
Indeed, according to the erasure rules \verb;view1; erases to the term
\verb;λ x. x;. We introduce a type \verb;Id X Y;, which is the set of all
functions from \verb;X; to \verb;Y; that erase to the identity function
(\verb;λ x. x;):
\begin{alltt}
id ◂ ∀ X : ★. X ➔ X = Λ X. λ x. x.

Id ◂ ★ ➔ ★ ➔ ★ = λ X : ★. λ Y : ★. Σ f : X ➔ Y. f ≃ id.
\end{alltt}

\paragraph{Introduction} The importance of the previously implemented combinator
\verb;subst; is that it allows to introduce an identity function
\verb;Id X Y; from any extensional identity \verb;f : X ➔ Y; (i.e.,
\verb;f x ≃ x; for any \verb;x;):
\begin{alltt}
intrId ◂ ∀ X Y : ★. Π f : X ➔ Y. (Π x : X. f x ≃ x) ➔ Id X Y 
 = Λ X. Λ Y. λ f. λ prf. pair (λ x. subst x -(f x) -(prf x)) β.
\end{alltt}

\paragraph{Elmination} Given an identity function \verb;c : Id X Y; and a value \verb;x : X;
we can apply the identity function \verb;c; to \verb;x; so that
\verb;elimId -c x; has type \verb;Y;:
\begin{alltt}
elimId ◂ ∀ X Y : ★. ∀ c : Id · X · Y. X ➔ Y =
  = Λ X. Λ Y. Λ c. λ x. subst x -(π₁ c x) -(ρ (π₂ c) - β).
\end{alltt}
The subterm \verb;ρ (π₂ c) - β; proves \verb;π₁ c x ≃ x;, where \verb;πᵢ; is the \verb;i;-th projections from a dependent product. Observe that
\verb;elimId; itself erases to \verb;λ x. x;, hence
\verb;elimId -c x ≃ x; by beta-reduction.
In other words, an identity
function \verb;Id X Y; allows \verb;x : X; to be seen as having types
\verb;X; and \verb;Y; at the same time.

\subsection{Identity Mapping}
A scheme \verb;F : ★ ➔ ★; is a functor if it comes equipped with a function
\verb;fmap; that satisfies the identity and composition laws:
\begin{alltt}
Functor ◂ (★ ➔ ★) ➔ ★ = λ F : ★ ➔ ★. 
  Σ fmap : ∀ X : ★. ∀ Y : ★. (X ➔ Y) ➔ F X ➔ F Y.
  IdentityLaw fmap × CompositionLaw fmap.
\end{alltt}
However, it is simple to define a covariant scheme for which the function
\verb;fmap; cannot be implemented (below, \verb;x₁ ≠ x₂; is shorthand for \verb;x₁ ≃ x₂ ➔ Empty;):
\begin{alltt}
UneqPair ◂ ★ ➔ ★ = λ X : ★. Σ x₁ : X. Σ x₂ : X. x₁ ≠ x₂.
\end{alltt}

We introduce schemes with \emph{identity mappings} as a new class of
parameterized covariant schemes. An identity mapping is a lifting
of identity functions:
\begin{alltt}
IdMapping ◂ (★ ➔ ★) ➔ ★ = λ F : ★ ➔ ★. 
  ∀ X Y : ★. Id X Y ➔ Id (F X) (F Y).
\end{alltt}
Intuitively, \verb;IdMapping F; is similar to \verb;fmap; of functors,
but it needs to be defined only on identity functions.  The identity law is
expressed as a requirement that identity function \verb;Id X Y; is
mapped to identity function \verb;Id (F X) (F Y);.

Clearly, every functor induces an identity mapping (by the application of
\verb;intrId; to \verb;fmap; and its identity law):
\begin{alltt}
fm2im ◂ ∀ F : ★ ➔ ★. Functor · F ➔ IdMapping · F = <..>
\end{alltt}
However, \verb;UneqPair; is an example of scheme which is not a
functor, but has an identity mapping (see example in
Section~\ref{unbalanced-trees}).

In the rest of the paper we show that every identity mapping
\verb;IdMapping F; induces an inductive datatype which is a least
fixed point of \verb;F;. Additionally, we generically derive an induction
principle and implement a constant-time
destructor for these datatypes.

\section{Inductive Datatypes from Identity Mappings}
In our previous paper, we used Cedille to show how to generically
derive an induction principle for Mendler-style lambda-encoded
datatypes that arise as least fixed points of
functors~\cite{firsov18}. In this section, we revisit this derivation
to show that it is possible to relax functoriality constraints and
only assume that the underlying signature scheme is accompanied by
an identity mapping.

\subsection{Basics of Mendler-Style Encoding}
\label{mendler-encoding}
In this section, we investigate the standard definitions of
Mendler-style F-algebras that are well-defined for any unrestricted
scheme \verb;F : ★ ➔ ★;. To reduce the notational clutter, we assume
that \verb;F : ★ ➔ ★; is a global (module) parameter:
\begin{alltt}
module _ (F : ★ ➔ ★)
\end{alltt}

In the abstract setting of category theory, a Mendler-style F-algebra is a
pair $(X,\Phi)$ where $X$ is an object (i.e., the \emph{carrier}) in
$\mathcal{C}$ and $\Phi : \mathcal{C}(-,X) → \mathcal{C}(F\ -, X)$ is
a natural transformation. In the concrete setting of Cedille, objects
are types, arrows are functions, and natural transformations are polymorphic
functions. Therefore, Mendler-style F-algebras are defined as follows:
\begin{alltt}
AlgM ◂ ★ ➔ ★ = λ X : ★. ∀ R : ★. (R ➔ X) ➔ F · R ➔ X.
\end{alltt}
Uustalu and Vene showed that initial Mendler-style F-algebras offer an
alternative categorical model of inductive
datatypes~\cite{uustalu99}. The carrier of an initial F-algebra is an
inductive datatype that is a least fixed point of \verb;F;. It is
known that if \verb;F; is a positive scheme then the least fixed point
of it may be implemented in terms of universal
quantification~\cite{wadler90}:
\begin{alltt}
FixM ◂ ★ = ∀ X : ★. AlgM X ➔ X.

foldM ◂ ∀ X : ★. AlgM X ➔ FixM ➔ X = Λ X. λ alg. λ x. x alg.
\end{alltt}
In essence, this definition identifies inductive datatypes with
iterators and every function on \verb;FixM; is to be computed by
iteration.

The natural transformation of the initial Mendler-style F-algebra
denotes the collection of constructors of its
carrier~\cite{uustalu99}.  In our setting, the initial Mendler-style
F-algebra \verb;AlgM FixM; is
not defineable because \verb;F; is not a
functor~\cite{firsov18}.  Instead, we express the collection of
constructors of datatype \verb;FixM; as a conventional F-algebra
\verb;F FixM ➔ FixM;:
\begin{alltt}
inFixM ◂ F FixM ➔ FixM = λ x. Λ X. λ alg. alg (fold alg) x.
\end{alltt}
The function \verb;inFixM; is of crucial importance because it expresses
constructors of \verb;FixM; without requirements of functoriality on
\verb;F : ★ ➔ ★;.

It is provably impossible to define the mutual inverse of
\verb;inFixM; (destructor of \verb;FixM;) without introducing
additional constraints on \verb;F;. Assume the existence of function
\verb;outFixM; (which need not be an inverse of \verb;inFixM;),
typed as follows:
\begin{alltt}
outFixM ◂ ∀ F : ★ ➔ ★. FixM F ➔ F (FixM F) = <..>
\end{alltt}
Next, recall that in the impredicative setting the empty type is
encoded as \verb;∀ X : ★. X; (its inhabitant implies any
equation). Then, we instantiate \verb;F; with the negative polymorphic
scheme \verb;NegF X := ∀ Y: ★. X ➔ Y;, and exploit the function
\verb;outFixM; to construct a witness of the empty type:
\begin{alltt}
T ◂ ★ = FixM · NegF.

ty ◂ ∀ Y : ★. T ➔ Y = Λ Y. λ t. outFixM · NegF t · Y t.

t ◂ T = ty · T (inFixM · NegF ty).

unsound ◂ ∀ X : ★. X = Λ X. ty · X t.
\end{alltt}
Therefore, the existance of function \verb;outFixM; contradicts the
consistency of Cedille. Hence, the inverse of \verb;inFixM; can exist
only for some restricted class of schemes \verb;F : ★ ➔ ★;.

\subsection{Inductive Subset}
From this point forward we assume that the scheme
\verb;F; is also accompanied by an identity mapping \verb;imap;:
\begin{alltt}
module _ (F : ★ ➔ ★)(imap : IdMapping · F)
\end{alltt}

In our previous work we assumed that \verb;F; is a functor and
showed how to specify the ``inductive'' subset of the type
\verb;FixM F;. Then, we generically derived induction for this
subset. In this section, we update the steps of our previous
work to account for \verb;F : ★ ➔ ★; not being a
functor.

The dependent intersection type \verb;ι x : X. Y x; can be understood as a
subset of \verb;X; defined by a predicate \verb;Y;. However, to
construct the value of this type we must provide \verb;x : X; and a
proof \verb;p : Y x; so that \verb;x; and \verb;p; are provably equal
(\verb;x ≃ p;). Hence, to align with this constraint we use implicit
products to express inductivity of \verb;FixM; as its
``dependently-typed'' version. Recall that \verb;FixM; is defined in
terms of Mendler-style F-algebras:
\begin{alltt}
AlgM ◂ ★ ➔ ★ = λ X : ★. ∀ R : ★. (R ➔ X) ➔ F · R ➔ X.
\end{alltt}
In our previous work, we introduced the \emph{Q-proof F-algebras} as a
``dependently-typed'' counterpart of \verb;AlgM;. The value of type
\verb;PrfAlgM X Q alg; should be understood as an inductive proof that
predicate \verb;Q; holds for every \verb;X; where \verb;X; is a least
fixed point of \verb;F; and \verb;alg : F X ➔ X; is a collection of
constructors of \verb;X;.
\begin{alltt}
PrfAlgM ◂ Π X : ★. (X ➔ ★) ➔ (F · X ➔ X) ➔ ★ 
 = λ X : ★. λ Q : X ➔ ★. λ alg : F · X ➔ X. 
 ∀ R : ★. ∀ c : Id · R · X. (Π r : R. Q (elimId -c r)) ➔
 Π fr : F · R. Q (alg (elimId -(imap c) fr)).
\end{alltt}
Mendler-style F-algebras (\verb;AlgM;) allow recursive calls
to be explicitly stated by providing arguments \verb;R ➔ X; and
\verb;F R;, where the polymorphically quantified type \verb;R; ensures
termination.  Similarly, Q-proof F-algebras allow
the inductive hypotheses to be explicitly stated for every \verb;R; by
providing an implicit
identity function \verb;c : Id R X;, and a dependent function of type
\verb;Π r : R. Q (elimId -c r); (recall that \verb;elimId -c r;
reduces to \verb;r; and has type \verb;X;). Given the inductive hypothesis for
every \verb;R;, the proof algebra must conclude that the predicate
\verb;Q; holds for every \verb;X;, which is produced by constructors \verb;alg;
from any given \verb;F R; that has been ``casted'' to \verb;F X;.

Next, recall that \verb;FixM; is defined as a function from \verb;AlgM X; to 
\verb;X; for every \verb;X;.
\begin{alltt}
FixM ◂ ★ =  ∀ X : ★. AlgM X ➔ X.
\end{alltt}
To retain the analogy of definitions, we express the inductivity of
value \verb;x : FixM; as a dependent function from a \emph{Q}-proof
\emph{F}-algebra to \verb;Q x;.
\begin{alltt}
IsIndFixM ◂ FixM ➔ ★ = λ x : FixM. 
 ∀ Q : FixM ➔ ★. PrfAlgM FixM Q inFixM ➔ Q x.
\end{alltt}
Now, we employ intersection types to define a type \verb;FixIndM; as a
subset of \verb;FixM; carved out by the ``inductivity'' predicate
\verb;IsIndFixM;:
\begin{alltt}
FixIndM ◂ ★ = ι x : FixM. IsIndFixM x.
\end{alltt}
Finally, we must explain how to construct the values of this type. As in
the case of \verb;FixM;, the set of constructors of \verb;FixIndM; is
expressed by a conventional F-algebra
\verb;F FixIndM ➔ FixIndM;.
The implementation is divided into three steps:

First, we define a function from \verb;F FixIndM; to \verb;FixM;:
\begin{alltt}
tc1 ◂ F FixIndM ➔ FixM = λ x. 
 let c ◂ Id (F FixIndM) (F FixM) = imap (intrId (λ x. x.1) β) in 
 inFixM (elimId -c x).
\end{alltt}
The implementation simply ``casts'' its argument to \verb;F FixM; and
then applies the previously implemented constructor of \verb;FixM;
(\verb;inFixM;).  Because \verb;elimId -c x; reduces to \verb;x;, the
erasure of \verb;tc1; is the same as the erasure of \verb;inFixM;
which is a term \verb;λ x. λ q. q (λ r. r q) x;.

Second, we show that the same lambda term could also be typed as a
proof that every \verb;tc1 x; is inductive:
\begin{alltt}
tc2 ◂ Π x : F · FixIndM. IsIndFixM (tc1 x)
 = λ x. (Λ Q. λ q. (q -(intrId  (λ x. x.1) β) (λ r. r.2 q) x)).
\end{alltt}
Indeed, functions \verb;tc1; and \verb;tc2; are represented by the
same pure lambda term. 

Finally, given any value \verb;x : F FixInd; we can intersect
\verb;tc1 x; and the proof of its inductivity \verb;tc2 x; to
construct an element of an inductive subset \verb;FixIndM;:
\begin{alltt}
inFixIndM ◂ F · FixIndM ➔ FixIndM = λ x. [ tc1 x, tc2 x \{ β \} ].
\end{alltt}
Recall that erasure of intersection \verb;[ x, y {p} ]; equals
the erasure of \verb;x;. Therefore, functions \verb;tc1;, \verb;tc2;,
and \verb;inFixIndM; all erase to the same pure lambda term. In other
words, in Cedille the term \verb;λ x. λ q. q (λ r. r q) x; can be
extrinsically typed as any of these functions.

\subsection{Induction Principle}
We start by explaining why we need to derive induction
for \verb;FixIndM;, even though it is definitionally an inductive subset of
\verb;FixM;. Indeed, every value \verb;x : FixIndM; can be ``viewed''
as a proof of its own inductivity. More precisely, the term \verb;x.2;
is a proof of the inductivity of \verb;x.1;. Moreover, the equational
theory of CDLE gives us the following equalities \verb;x.1 ≃ x ≃ x.2;
(due to the rules of erasure). But recall that the inductivity proof
provided by the second view \verb;x.2; is typed as follows:
\begin{alltt}
∀ Q : Fix  ➔ ★. PrfAlgM FixM · Q inFixM ➔ Q x.1 
\end{alltt}
Note that \verb;Q; is a predicate on \verb;FixM; and not
\verb;FixIndM;! This form of inductivity does not allow
properties specified directly for \verb;FixIndM; to be proven.

Therefore, our goal is to prove that every \verb;x : FixIndM; is
inductive in its own right. We phrase this in terms of proof-algebras
parameterized by \verb;FixIndM;, a predicate on \verb;FixIndM;, and its
constructors (\verb;inFixIndM;):
\begin{alltt}
∀ Q : FixIndM ➔ ★. PrfAlgM FixIndM · Q inFixIndM ➔ Q x.
\end{alltt}
In our previous work, we already made an observation that the derivation
of induction for Mendler-style encodings relies only on the identity
law of functors~\cite{firsov18}. Therefore, the current setting
only requires minor adjustments of our previous proof. For the sake of
completeness, we present a main idea of this derivation.

The key insight is that we can convert predicates on \verb;FixIndM; to
logically equivalent predicates on \verb;FixM; by using heterogeneous
equality:
\begin{alltt}
Lift ◂ (FixInd ➔ ★) ➔ Fix ➔ ★
 = λ Q : FixInd ➔ ★.  λ y : Fix. Σ x : FixInd. x ≃ y × Q x.

eqv1 ◂ Π x: FixIndM. ∀ Q: FixIndM ➔ ★. Q x ➔ Lift · Q x.1 = <..>

eqv2 ◂ Π x: FixIndM. ∀ Q: FixIndM ➔ ★. Lift · Q x.1 ➔ Q x = <..>
\end{alltt}
These properties allow us to convert a \emph{Q}-proof algebra to a
proof algebra for a lifted predicate \verb;Lift Q;, and then derive the
generic induction principle:
\begin{alltt}
convIH ◂ ∀ Q : FixIndM ➔ ★. PrfAlgM FixIndM · Q inFixIndM 
 ➔ PrfAlgM FixM · (Lift · Q) inFixM = <..> 

induction ◂ ∀ Q: FixIndM ➔ ★. PrfAlgM FixIndM · Q inFixIndM ➔ 
 Π e: FixIndM. Q e = Λ Q. λ p. λ e. eqv2 e (e.2 (convIH p)).
\end{alltt}
Let \verb;Q; be a predicate on \verb;FixIndM; and \verb;p; be a
\emph{Q}-proof algebra: we show that \verb;Q; holds for any
\verb;e : FixIndM;.  Recall that every \verb;e : FixIndM; can be
viewed as a proof of inductivity of \verb;e.1; via
\verb;e.2 : IsIndFixM e.1;. We use this to get a proof of the lifted
predicate \verb;Lift Q e.1; from the proof algebra delivered by
\verb;convIH p;. Finally, we get \verb;Q e; by using \verb;eqv2;.

\section{Constant-Time Destructors}
An induction principle is needed to prove properties about programs, but
practical functional programming also requires constant-time destructors
(also called accessors) of inductive datatypes. Let us illustrate the
problem using the datatype of natural numbers. In Agda it is easy to implement the
predecessor function by pattern matching:
\begin{alltt}
pred : Nat ➔ Nat
pred zero = zero
pred (suc n) = n
\end{alltt}
The correctness of \verb;pred; trivially follows by beta-reduction:
\begin{alltt}
predProp : (n : Nat) ➔ pred (suc n) ≡ n
predProp n = refl
\end{alltt}
Let us switch to Cedille and observe that it is much less trivial to
implement the predecessor for the impredicative encoding of Peano
numerals. Here is the definition of Church encoded Peano naturals and
their constructors:
\begin{alltt}
cNat ◂ ★ = ∀ X : ★. (X ➔ X) ➔ X ➔ X.

zero ◂ cNat = Λ X. λ s. λ z. z.
suc  ◂ cNat ➔ cNat = λ n. Λ X. λ s. λ z. s (n s z).
\end{alltt}
Next, we implement the predecessor for \verb;cNat; which is due to
Kleene:
\begin{alltt}
zCase ◂ cNat × cNat = pair zero zero
sCase ◂ cNat × cNat ➔ cNat × cNat = λ n. pair (π₂ n) (suc (π₂ n)).

predK ◂ Nat ➔ Nat = λ n. π₁ (n sC zC).
\end{alltt}
The key to the Kleene predecessor is the function \verb;sCase;, which
ignores the first item of the input pair, moves the second natural to
the first position, and then applies the successor of the second
element within the second position. Hence, folding a natural number
\verb;n; with \verb;sCase; and \verb;zCase; produces a pair
\verb;(n-1, n);. In the end, \verb;predK n; projects the first element
of a pair.

Kleene predecessor runs in linear time. Also, \verb;predK (suc n); gets stuck after reducing to
\verb;π₁ (pair (π₂ (n sCase zCase))(suc (π₂ (n sCase zCase))));.
Hence, we must use induction to prove that \verb;predK (suc n);
computes to \verb;n;.

Furthermore, Parigot proved that any definition of predecessor for the
Church-style lambda-encoded numerals requires linear
time~\cite{parigot89}.


\subsection{Constant-Time Destructor for Mendler-Style Encoding}
In previous sections we defined a datatype \verb;FixIndM; for every
scheme \verb;F; that has an identity mapping. Then, we implemented the
constructors of the datatype as the function \verb;inFixIndM;,
and defined an induction
principle phrased in terms of this function. In this section, we
develop a mutual inverse of \verb;inFixIndM; that runs in constant
time. As a simple consequence, we prove
that \verb;FixIndM; is a least fixed point of \verb;F;.

Let us start by exploring the computational behaviour of the function
\verb;foldM;. The following property is a variation of the
\emph{cancellation law} for Mendler-style encoded
data~\cite{uustalu99}, and its proof is simply by beta-reduction.
\begin{alltt}
foldHom ◂ ∀ X : ★. Π x : F FixM. Π alg : AlgM X. 
  foldM alg (inFixM x) ≃ alg (foldM alg) x = Λ X. λ x. λ a. β.
\end{alltt}
In other words, folding the inductive value \verb;inFixM x; replaces
its outermost ``constructor'' \verb;inFixM; with partially applied F-algebra
\verb;alg (foldM alg);.

It is well-known that (computationally) induction can be reduced to
iteration (folding).
Therefore, we can state the cancellation law for
the induction rule in terms of proof algebras.
\begin{alltt}
indHom ◂ ∀ Q : FixIndM ➔ ★. Π alg : PrfAlgM FixIndM Q inFixIndM.
 Π x : F FixIndM. Π c : Id FixIndM FixIndM.
 induction alg (inFixInd x) ≃ alg -c (induction alg) x 
 = Λ Q. λ p. λ x. β.
\end{alltt}
Most importantly, is that the proof of \verb;indHom; is by reflexivity
(\verb;β;), which ensures that the left-hand side of equality
beta-reduces to the right-hand side in a constant number of
beta-reductions.

Next, we implement a proof algebra for the constant predicate
\verb;λ _. F FixIndM;.
\begin{alltt}
outAlgM ◂ PrfAlgM FixIndM (λ _. F FixIndM) inFixIndM
 = Λ R. Λ c. λ f. λ y.  elimId -(imap c) y.
\end{alltt}
The identity mapping of \verb;F; lifts the identity function
\verb;c : Id R X; to an identity function \verb;Id (F R) (F FixIndM);,
which is then applied to the argument \verb;y : F R; to get the desired
value of \verb;F FixIndM;.

The proof algebra \verb;outAlgM; induces the constant-time inverse of
\verb;inFixIndM;:
\begin{alltt}
outFixIndM ◂ FixInd ➔ F FixInd = induction outAlgM.
\end{alltt}
Definitionally, \verb;outFixIndM (inFixIndM x); is
\verb;induction outAlgM (inFixInd x);, which reduces to
\verb;outAlgM -c (induction outAlgM) x; in a constant number of steps
(\verb;indHom;).  Because \verb;outAlgM -c; erases to
\verb;λ f. λ y. y;, it follows that \verb;outFixIndM; computes an
inverse of \verb;inFixIndM; in a constant number of beta-reductions:
\begin{alltt}
lambek1 ◂ Π x: F · FixInd. outFixIndM (inFixIndM x) ≃ x = λ x. β.
\end{alltt}
Furthermore, we show that \verb;outFixIndM; is a post-inverse:
\begin{alltt}
lambek2 ◂ Π x: FixIndM. inFixIndM (outFixIndM x) ≃ x
 = λ x. induction (Λ R. Λ c. λ ih. λ fr. β) x.
\end{alltt}
This direction requires us to invoke induction to ``pattern match'' on
the argument value to get \verb;x := inFixIndM y; for some value
\verb;y; of type \verb;F FixIndM;. Then,
\verb;inFixIndM (outFixIndM (inFixIndM y)) ≃ inFixIndM y;
because the inner
term \verb;outFixIndM (inFixIndM y); is just \verb;y; by beta
reduction (\verb;lambek1;).


\section{Examples}
In this section, we demonstrate the utility of our derivations on three
examples. First, we present a detailed implementation of natural
numbers with a constant-time predecessor function. Second, we show examples
of infinitary datatypes. Finally, we give an example of a datatype
arising as a least fixed point of a scheme that is not a functor, but
has an identity mapping.

\subsection{Natural Numbers with Constant-Time Predecessor}
Natural numbers arise as a least fixed point of the functor \verb;NF;:
\begin{alltt}
NF ◂ ★ ➔ ★ = λ X : ★. Unit + X.

nfmap ◂ Functor NF = <..>
\end{alltt}
Since every functor has an identity mapping then we use our framework
to define natural numbers as shown below:
\begin{alltt}
nfimap ◂ IdMapping · NF = fm2im nfmap.

Nat ◂ ★ = FixInd · NF nfimap.

zero ◂ Nat = inFixIndM (in1 unit).
suc  ◂ Nat ➔ Nat = λ n. inFixIndM (in2 n).
\end{alltt}
If injections \verb;in1; and \verb;in2; erase to
\verb;λ a. λ i. λ j. i a; and \verb;λ a. λ i. λ j. j a;,
respectively, then the natural number constructors have the following
erasures:
\begin{alltt}
zero  ≃ λ alg. (alg (λ f. (f alg)) (λ i. λ j. (i (λ x. x))))
suc n ≃ λ alg. (alg (λ f. (f alg)) (λ i. λ j. (j n)))
\end{alltt}
Intuitively, Mendler-style numerals have a constant-time predecessor
because every natural number \verb;suc n; contains the previous natural
\verb;n; as its direct subpart (which is not true for Church
encoding).

We implement the predecessor for \verb;Nat; in terms of the generic
constant-time destructor \verb;outFixIndM;:
\begin{alltt}
pred ◂ Nat ➔ Nat = λ n. case (outFixIndM n) (λ _. zero) (λ m. m).
\end{alltt}
Because elimination of disjoint sums (via \verb;case;) and \verb;outFixIndM;
are both constant-time operations, \verb;pred; is also a constant-time
function and its correctness is immediate (i.e., by beta-reduction):
\begin{alltt}
predSuc ◂ Π n : Nat. pred (suc n) ≃ n = λ n. β.
\end{alltt}

We also show that the usual ``flat'' induction principle can be
derived from our generic induction principle (\verb;induction;) by dependent
elimination of \verb;NF;:
\begin{alltt}
indNat ◂ ∀ P : Nat ➔ ★. (Π n : Nat. P n ➔ P (suc n)) ➔ P zero
 ➔ Π n : Nat. P n = Λ P. λ s. λ z. λ n. induction · P 
 (Λ R. Λ c. λ ih. λ v. case v (λ u. ρ (etaUnit u) - z) 
                              (λ r. s (elimId -c r) (ih r))) n.
\end{alltt}

\subsection{Infinitary Trees}
\label{sec:inftree}
In Agda, we can give the following inductive definition of
infinitary trees:
\begin{alltt}
data ITree : Set where
  node : (Nat ➔ Unit + ITree) ➔ ITree
\end{alltt}
\verb;ITree; is a least fixed point of functor
\verb;IF X := Nat ➔ Unit + X;. In Cedille, we can implement a
functorial function lifting for \verb;IF;:
\begin{alltt}
itfmap ◂ ∀ X Y : ★. (X ➔ Y) ➔ IF X ➔ IF Y
 = λ f. λ t. λ n. case (t n) (λ u. in1 u) (λ x. in2 (f x)).
\end{alltt}
To our best knowledge, it is impossible to prove that \verb;itfmap;
satisfies the functorial laws without functional extensionality (which
is unavailable in Cedille). However, it is possible to implement an
\emph{identity mapping} for the scheme \verb;IF;:
\begin{alltt}
itimap ◂ IdMapping · IF
 = Λ c. pair (λ x. λ n. elimId -(nfimap -c) (x n)) β.
\end{alltt}
The first element of a pair erases to \verb;λ x. λ n. x n;, which is
\verb;λ x. x; by the eta law\footnote{Our analysis of CDLE up to now
  has included only $\beta$-equality. It is known that $\eta$ can
  cause problems for intrinsic type theories due to non-confluence of
  $\beta\eta$-reduction on ill-typed terms (cf.~\cite{geuvers93}).
  But for extrinsic typing, we can use confluence of
  $\beta\eta$-reduction on pure lambda terms, and thus we hope that
  adding $\eta$ does cause problems. The use of $\eta$ is confined to
  the example of infinitary trees only.}. Now, since we showed that
\verb;IF; has an identity mapping then our generic development
induces the datatype \verb;ITree; with its constructor, destructor, and
induction principle.
\begin{alltt}
ITree ◂ ★ = FixIndM IF itimap.

inode ◂ (Nat ➔ Unit + ITree) ➔ ITree = λ f. inFixIndM f.
\end{alltt}
The specialized induction is phrased in terms of ``empty tree''
\verb;iempty; which acts as a base case (\verb;projR;
``projects'' a tree from disjoint sum or returns \verb;iempty;):
\begin{alltt}
iempty ◂ ITree = inode (λ _. in1 unit).

indITree ◂ ∀ P : ITree ➔ ★. P iempty ➔ 
 (Π f: Nat ➔ Unit + ITree. (Π n : Nat. P (projR (f n))
   ➔ P (inode f)) ➔ Π t: ITree. P t = <..>
\end{alltt}

Next, let us look at another variant of infinitary datatypes in Agda:
\begin{alltt}
data PTree : Set where
  pnode : ((PTree ➔ Bool) ➔ Unit + PTree) ➔ PTree
\end{alltt}
This definition will be rejected by Agda (and Coq) since it arises as
a least fixed point of the scheme
\verb;PF X := Unit + ((X ➔ Bool) ➔ X) ➔ X;, which is positive but not
strictly positive. The definition is rejected because it is currently
unclear if non-strict definitions are sound in Agda. For the Coq setting, there is a
proof by Coquand and Paulin that non-strict positivity combined with an
impredicative universe and a predicative universe hieararchy leads to
inconsistency~\cite{coquand90}.  In Cedille, we can implement an
identity mapping for the scheme \verb;PF; in a similar fashion as the
previously discussed \verb;UF;. Hence, the datatype induced
by \verb;PF; exists in the type theory of Cedille.

\subsection{Unbalanced Trees}
\label{unbalanced-trees}
Consider the following definition of ``unbalanced'' binary trees in Agda:
\begin{alltt}
data UTree : Set where
  leaf : Bool → UTree
  node : (b₁ : UTree) ➔ (b₂ : UTree) ➔ b₁ ≠ b₂ ➔ UTree
\end{alltt}
The datatype \verb;UTree; arises as a least fixed point of the
following scheme:
\begin{alltt} 
UF ◂ ★ ➔ ★ = λ X : ★. Bool + (Σ x₁ : X. Σ x₂ : X. x₁ ≠ x₂).
\end{alltt}
Because the elements \verb;x₁; and \verb;x₂; must be different,
lifting an arbitrary function \verb;X ➔ Y; to
\verb;UF X ➔ UF Y; is impossible. Hence, the scheme \verb;UF; is not a
functor.

However, we can show that \verb;UF; has an identity mapping. We start by
producing a function \verb;UF X → UF Y; from an identity \verb;Id X Y;:
\begin{alltt}
uimap' ◂ ∀ X Y : ★. ∀ i : Id · X · Y. UF X ➔ UF Y = Λ i. λ uf. 
 case uf (λ u. in1 u) 
         (λ u. in2 (pair (elimId -i (π₁ u))
                   (pair (elimId -i (π₁ (π₂ u))) (π₂ (π₂ u)))))
\end{alltt}
We prove that \verb;uimap' -i; is extensionally an identity function:
\begin{alltt}
uimP ◂ ∀ X Y: ★. ∀ i: Id X Y. Π u: UF · X. uimap' -i u ≃ u = <..>
\end{alltt}
This is enough to derive an identity mapping for \verb;UF; by using
the previously implemented combinator \verb;intrId;:
\begin{alltt}
uimap ◂ IdMapping · UF = intrId uimap' uimP.
\end{alltt}
Therefore, we conclude that the datatype of unbalanced trees exists in
Cedille and can be defined as a least fixed point of scheme \verb;UF;:
\begin{alltt}
UTree ◂ ★ = FixIndM UF uimap.
\end{alltt}
The specialized constructors, induction principle, and a destructor
function for \verb;UTree; are easily derived from their generic
counterparts (\verb;inFixIndM;, \verb;induction;, \verb;outFixIndM;).

\section{Related Work}
Pfenning and Paulin-Mohring show how to model inductive datatypes
using impredicative encodings in the Calculus of
Constructions (CC)~\cite{pfenning89}. Because induction is not provable in the
CC, the induction principles are generated and added as
axioms. This approach was adopted by initial versions of the Coq proof
assistant, but later Coq switched to the Calculus of Inductive
Constructions (CIC), which has built-in inductive datatypes.

Delaware et al. derived induction for impredicative lambda-encodings
in Coq as a part of their framework for modular definitions and proofs
(using the à la carte technique~\cite{swierstra08}). They showed that
a value \verb;v : Fix F; is inductive if it is accompanied by a proof
of the universal property of folds~\cite{delaware13}.

Similarly, Torrini introduced the \emph{predicatisation} technique,
reducing dependent induction to proofs that only rely on non-dependent
Mendler induction (by requiring the inductive argument to satisfy
extra predicatisation hypothesis)~\cite{torrini16}.

Traytel et al. present a framework for constructing (co)datatypes in
HOL~\cite{Traytel+12,Biendarra+17}. The main ingredient is a notion of
a bounded natural functor (BNF), or a binary functor with additional
structure. BNFs are closed under composition and fixed points, which
enables support for both mutual and nested (co)recursion with mixed
combinations of datatypes and codatatypes. The authors developed a package
that can generate (co)datatypes with their associated
proof-principles from user specifications (including custom
bounded natural functors). In constrast, our approach provides a single generic
derivation of induction within the theory of Cedille, but does not address
codatatypes. It would be interesting to further investigate the exact
relationship between schemes with identity mappings and BNFs.

Church encodings are typeable in System F and represent
datatypes as their own iterators. Parigot proved that
the lower bound of the predecessor function for Church
numerals has linear time complexity~\cite{parigot89}.

Parigot designed an impredicative lambda-encoding that is typeable in
System F$_\omega$ with positive-recursive type definitions. The
encoding identifies datatypes with their own recursors, allowing
constant time destructors to be defined, but the drawback is that the
representation of a natural number $n$ is exponential in
the call-by-value setting~\cite{parigot88}.

The Stump-Fu encoding is also typeable in System F$_\omega$ with
positive-recursive type definitions. It improves upon the Parigot
representation by requiring only quadratic space, and it also supports
constant-time destructors~\cite{stump16}.

\section{Conclusions and Future Work}
In this work, we showed that the Calculus of Dependent Lambda
Eliminations is a compact pure type theory that allows a general class
of Mendler-style lambda-encoded inductive datatypes to be defined as
least fixed points of schemes with identity mappings. We also gave a
generic derivation of induction and implemented a constant-time
destructor for these datatypes. We used our development to give the
first example (to the best of our knowledge) of lambda-encoded natural
numbers with: provable induction, a constant-time predecessor
function, and a linear size (in the numeral $n$) term
representation. Our formal development is around 700 lines of Cedille
code.

For future work, we plan to explore coinductive definitions and to use the
categorical model of Mendler-style datatypes to investigate
histomorphisms and inductive-recursive datatypes in
Cedille~\cite{uustalu99}.


\subsubsection*{Acknowledgments.} We gratefully acknowledge NSF support under award 1524519, and DoD
support under award FA9550-16-1-0082 (MURI program).

\bibliography{paper}

\end{document}